\documentclass{aastex6}

\usepackage{rotating}
\fullcollaborationName{MASTER Global Robotic Net}

\begin{document}

\title{MASTER optical polarization variability detection in the  Microquasar V404 Cyg/GS2023+33}


\author{Vladimir M. Lipunov \altaffilmark{1,2},E.Gorbovskoy\altaffilmark{2},V. Kornilov\altaffilmark{1,2}, V.Krushinskiy\altaffilmark{3}, D.
Vlasenko\altaffilmark{1,2}, N.Tiurina \altaffilmark{2}, P.Balanutsa\altaffilmark{2}, A. Kuznetsov\altaffilmark{2}, N. Budnev\altaffilmark{4}, O. Gress\altaffilmark{4}, A. Tlatov\altaffilmark{5},
R.Rebolo Lopez \altaffilmark{6}, M. Serra-Ricart\altaffilmark{6},  D.A.H. Buckley\altaffilmark{7}, G. Israelyan\altaffilmark{6},   N. Lodieu\altaffilmark{6}, K.Ivanov\altaffilmark{4}, S.Yazev\altaffilmark{4}, Yu.Sergienko\altaffilmark{8}, A.Gabovich\altaffilmark{8}, V.Yurkov\altaffilmark{8}, H. Levato\altaffilmark{9}, C.Saffe\altaffilmark{9}, R.Podesta \altaffilmark{10}, C. Mallamaci\altaffilmark{10}, C.Lopez\altaffilmark{10}}
\affil{M.V.Lomonosov Moscow State University, Physics Department, Leninskie gory, GSP-1, Moscow, 119991, Russia }
\affil{M.V.Lomonosov Moscow State University, Sternberg Astronomical Institute, Universitetsky~pr.,~13,~Moscow,
119234, Russia}
\affil{ Kourovka Astronomical Observatory, Ural Federal University, Lenin ave. 51, Ekaterinburg 620000, Russia}
\affil{Applied Physics Institute, Irkutsk State University, 20, Gagarin blvd,664003, Irkutsk, Russia}
\affil{Kislovodsk Solar Station of the Main (Pulkovo) Observatory RAS, P.O.Box 45, ul. Gagarina 100, Kislovodsk
357700, Russia}
\affil{Instituto de Astrofísica de Canarias, C/Via Láctea, s/n E38205, La Laguna, Tenerife, Spain }
\affil{South African Astronomical Observatory, P.O. Box 9, Observatory 7935, Cape Town, South Africa}
\affil{Blagoveschensk State Pedagogical University, Lenin str., 104, Amur Region, Blagoveschensk 675000, Russia}
\affil{Instituto de Ciencias Astronomicas, de la Tierra y del Espacio (ICATE), Av.Espana Sur 1512, J5402DSP, San
Juan, Argentina}
\affil{Observatorio Astronomico Felix Aguilar (OAFA), Sede Central:  Avda Benavides s/n,  Rivadavia, El Leonsito,
Argentina}



\begin{abstract}
On 2015 June 15 the Swift space observatory discovered that the Galactic black hole candidate V404 Cyg was undergoing another active X-ray phase, after 25 years of inactivity ~\citep{7}. Twelve telescopes of the MASTER Global Robotic Net located at six sites across four continents were the first ground based observatories to start optical monitoring of the microquasar after its gamma-ray wakeup at 18h 34m 09s U.T. on 2015 June 15 \citep{9}.
 In this paper we report, for the first time, the discovery of variable optical linear polarization, changing by 4-6\% over a timescale of $\sim$1 h, on two different epochs. We can conclude that the additional variable polarization arisies from the relativistic jet generated by the black hole in V404Cyg. The polarization variability correlates with optical brightness changes, increasing when the flux decreases.
\end{abstract}

\keywords{black holes, V404 Cyg ---
polarization  --- optical jet --- x-ray nova --- relativistic jet}



\section{Introduction} \label{sec:intro}
The X-ray nova GS 2023+338 was discovered on 1989 May 22 by the Ginga X-ray satellite ~\citep{1}.
Soon after the object was identified with the variable star V404 Cyg, which had been seen to brighten by seven magnitudes
in the optical in 1938 ~\citep{2,3}. It is now well established ~\citep{4} that V404 Cyg is a binary system consisting of a
giant secondary star, older and less massive than the Sun
and a primary star which is an accreting black hole candidate \citep{BestBH,casares}. The secondary (optical companion) is believed to undergo episodic
enhanced mass loss every few decades, losing part of its mass to the black hole. This matter emits
X-ray radiation while heating up to several hundred million degrees in the accretion disk. The detection of
nonthermal radio and soft X-ray radiation ~\citep{6}, typically found in quasars, that are millions of times more
massive and with powerful black holes residing in nuclei of active galaxies (i.e. AGN), led  researchers to call this black hole a
microquasar, ejecting relativistic particles in the form of jets. In this paper we report the detection,  for the first
time, of variable optical polarization arising in the relativistic jet generated by the black hole.

On 2015 June 15 the Swift space observatory found that one of the most securely identified Galactic stellar mass black hole candidates, the binary system V404 Cyg,  began undergoing another active X-ray phase after 25 years of inactivity ~\citep{7}.

The twin optical telescope, MASTER-Tunka of the  MASTER Global Robotic Net ~\citep{8}, located near Lake Baikal (Lomonosov Moscow State University and Irkutsk State University Tunka astrophysical center), was the first ground-based observatory to point to V404 Cyg after the Swift alert. This was 22 sec after the notice time \citep{9} of the burst alert, via space communication system (i.e. socket messages \textbf{GCN: The Gamma-ray Coordinates Network alert \citep{Barthelmy94,Barthelmy95}}) 
and after 150 seconds we detected a bright optical flare ~\citep{9}, reaching magnitude 14, with the star brightening by a factor of 2.5  in a mere half-hour. In the next few weeks six MASTER twin robotic telescopes performed
about 20 pointings. Dozens of telescopes worldwide subsequently observed the object at various wavelengths
(~\citet{4,10,11,12,13,14,15,16}), but MASTER is the first to report on the discovery of variable optical linear polarization arising in the relativistic jet generated by the black hole.

\section{MASTER Global Robotic Network observation}

The MASTER robotic telescopes\footnote{\url{http://observ.pereplet.ru}} are a network of identical  instruments deployed over several continents
 and equipped with identical CCD cameras  capable of performing B,V,R$_J$,I (Johnson/Bessell)   photometric and linear polarimetric observations ~\citep{8,25,17}.  Each MASTER-II observatory has twin 0.40-m f/2.5 wide field reflectors (Hamilton design; see ~\cite{25}) and a prime focus 4k x 4k CCD camera, which provide a total  8 square-degree field and reaches to a white-light magnitude of 20-21 (in 180s exposure). In addition, the mounts for these two telescopes also have two very wide-field (VWF) optical cameras, with 800 square degrees field of view, reaching to magnitude of 15 \citep{8,25}.  The  observatories are located as follows:  operating in Russia, from east to west, the MASTER-Amur, MASTER-Tunka, MASTER-Ural, and MASTER-Kislovodsk; the MASTER-SAAO in South Africa;  the MASTER-IAC in the Canary Islands and MASTER-OAFA in Argentina.

The observations with MASTER-Net can be performed in three different modes: alert, survey and inspection. Alert mode is aimed at automated observations of rapid events, like Gamma Ray Bursts (GRBs) from SWIFT, Fermi, IPN, MAXI, INTEGRAL and gravitational wave or neutrino events. The primary goals of the MASTER Global Robotic Network are the rapid response to alerts (first, GRB alerts). The alert mode is triggered if a transient position has good accuracy (when the error-box size is less than  2$^\circ\times$2$^\circ$ MASTER FOV)  and is typically used to observe GRBs upon receiving notices from the Gamma-ray Coordinates Network\footnote{\url{http://gcn.gsfc.nasa.gov/}} . In this alert mode MASTER observes with parallel 0.4-m telescopes, each covering the same field,  and with polarizers \citep{Lipunov2016a}, pointing just after the notice time. Exposure times follow the relation $t_{exp} = (T_{start}-T_{0})/5$, where $T_{0}$ is the trigger time (UT), $T_{start}$ is the time of the beginning of exposure (UT). The exposure time is rounded to an integer with a step of 10 s and does not exceed 3 min.   All images are reduced  automatically in real-time mode via our  own software developed by the MASTER team over the past ten years. If we observe in survey or inspection mode, we usually observe in the white (unfiltered) light to increase the limiting magnitude. The corresponding internal photometric magnitudes can be described fairly well by the equation $W=0.8*R2+0.2*B2$, where $R2$ and $B2$ are the second epoch DSS red and blue magnitudes, respectively, adopted from USNO-B1.0 catalog ~\citep{monet,25,gor2012}.
 

 The technique of polarimetric measurements used on MASTER network telescopes is considered in detail in \citep{17}. It allows detecting the \textit{linear} polarization at 1-2\% level for objects brighter then 14mag. We stress that MASTER-II consists of twin wide field telescopes each with a 2$^\circ\times$2$^\circ$ field. It gives us the possibility to determine the polarization for thousands stars from 12 to 16 magnitudes simultaneously. To estimate an error of the object's polarimetric measurement we select a few tens of field stars with similar brigthness in the same frame and calculate the standard deviations of their polarizations, assuming that the stars have the same polarization produced by the ISM.
Therefore, we calculate the error of polarization measurements as a chance fluctuations of the background stars,
assuming that they have no variable polarization. 

\section {Polarimetry technique}
\label{section:obs}

The MASTER Net was designed with the objective to deliver polarization information as early as possible after GRB triggers. More than 100 observations of GRBs were made by the MASTER global robotic net. Optical emission was detected for $\sim$ 20 GRBs \citep{lipunov2007,gor2012,17}. The GRB\,100906A, GRB\,110422A, and GRB\,121011A events deserved attention because their optical observations were carried out during the prompt gamma-ray emission.

The technique of polarimetric measurements used on MASTER network telescopes is considered in detail in \citep{17} and although primarily designed for GRB followup, the same techniques can be used for polarimetry of other objects.

We only utilize linear polarizers for MASTER, so cannot determine the circular polarization of objects.

If we let $I_1$ and $I_2$ be the  signals on the CCD detector with orthogonal polarizing filters (for example oriented with their axes at $45^o$ and $135^o$), then a lower limit of the linear polarization is obtained from $P_{low}=(I_1-I_2)/(I_1+I_2)$.

To derive the degree of linear polarization, rather than a limit,  one would need to perform observations with polarizers positioned at three angles, for example, at $0^o$, $45^o$, and $90^o$ with respect to the reference direction. Since we
have  two fixed orthogonal filters at each of the MASTER-II telescopes, each telescope alone observes only a lower limit on polarization level --- $P_{low}$.

If the value of $P_{low}$ is less than measurement error $\sigma_\mathrm{p}$ , then formally we
have zero as a minimum estimate for the degree of linear polarization. Even with a 100-percent linearly polarized source, it is still possible, if the polarizers were so positioned that linear polarization position was inclined exactly by $45^o$ to
both of them, to register zero polarization for the object.

The two synchronous frames (taken with different cameras) used to measure $P_{low}$ are mutually calibrated so that the average $P_{low}$ for comparison stars would be $ Avg (P_{low}) = 0$. This is achieved at the stage of the photometric calibration. We use the same reference stars for frames in both polarizations from the USNO B-1 catalog (with no polarization measuremtns). This implies, for unpolarized sources, that $ Avg (I_{45} - I_{135}) = 0 $ and consequently $ Avg (P_{low})=0 $. If the reference stars are polarized (e.g. due to interstellar polarization, then they will show non-zero $P_{low}$.  For the purpose of this paper, the exact measurement of the linear polarization value is not the aim, but rather the purpose is to establish the polarization variability of V404Cyg.  Since the Galactic polarization is constant, it does not affect to the determination of polarization variability.

The MASTER polarization band is determined by the response curves of the CCD camera and a transmission curve of polarizing filter, which have been reviewed here \citep{17}.

For the determination of errors, we analyzed the distribution of $P_{low}$ of the field reference stars, depending on the magnitude. Then we selected only the stars with magnitudes in the interval $m_{V404} \pm 0.5 $, where $m_{V404}$ is a V404 magnitude on a given frame.  The dispersion of comparison stars $P_{low}$  in this interval are used to define the $1 - \sigma $ error of object $P_{low}$. We use the reference stars in a radius of $0.5^o$ for V404 Cyg.  All errors that are reported in this paper are  $1 \sigma$.

\section{MASTER V404 Cyg results}
In the direction of V404 Cyg, strong interstellar polarization is observed  due to scattering by
 Galactic dust, which is aligned with the Galactic magnetic field. Naturally this polarization is essentially constant, as
 noted in the very first reported observations of the new outburst of the microquasar
 \citep{18,19}. The magnitude  and position angle measured for V404 Cyg in ~\citep{19} was  found to be consistent with the interstellar linear polarization, at a level of about 8$\%$ and at a position angle of $\sim15^\circ$, respectively.

The MASTER-Net observations
were performed throughout 2015 June at the MASTER-Tunka and MASTER-Kislovodsk nodes. They reveal not only the constant interstellar polarization component during  the earliest epochs of optical observations, but also polarization variations at certain epochs in a number of cases. The full description of the data reduction process, polarization standards, photometry and polarization measurements are decribed in \citep{17}.


We observed in two orthogonally related polarizers we detect the value $Y=(I_{1}-I_{2})/(I_{1}+I_{2})$. It's evident that Y is the low limit of linear polarization of the object. But the Y change (time evolution of Y) authentically proves the  existence of an additional variable linear polarization, that is not connected with  the interstellar dust.

\textbf{As the Y value is measured in 2 polarizers with their fast axes orthogonal to each other, its variability is a result of the change of the E-vector, i.e. the degree of linear polarization and its position angle. The variability of flux in different polarizers can be  explained by the variation of the contribution of polarized flux to the total flux.  The degree of polarization increases when the unpolarized component of the optical flux of the system decreases relative to the contribution of the polarized flux. Therefore we expect the polarization to increase when the total optical brightness of system falls, say due to less X-ray irradiation of the secondary star. The polarized jet component then begins to dominate over the unpolarized reprocessed component, increasing the level of polarization.}


We found at least two cases of bona fide linear polarization variability. The first event lasted a
little more than an hour and was detected during alert observations on 2015 June 18. These observations were
triggered by an INTEGRAL Alert (GCN, trigger Number 7029) following the detection of an X-ray burst. The MASTER-Tunka
telescope automatically started observing the microquasar with the two parallel aligned telescope tubes, with mutually perpendicular polaroids,
positioned along the North-South and East-West directions, 42 seconds after the trigger. At 15:36:00 UT MASTER-Tunka
recorded the onset  of an abrupt factor of six decrease of optical flux (Fig.1), followed by its recovery at an
even higher level than the initial level. At the same time the polarization increased by $\sim4\%$, reaching
$\sim 12\%$, while it  was of about $\sim 8\%$ before and after this event (one of the axes of our polarizer is aligned practically parallel to the position angle of the interstellar polarization \citep{20}.

\begin{figure}
\figurenum{1}
\plotone{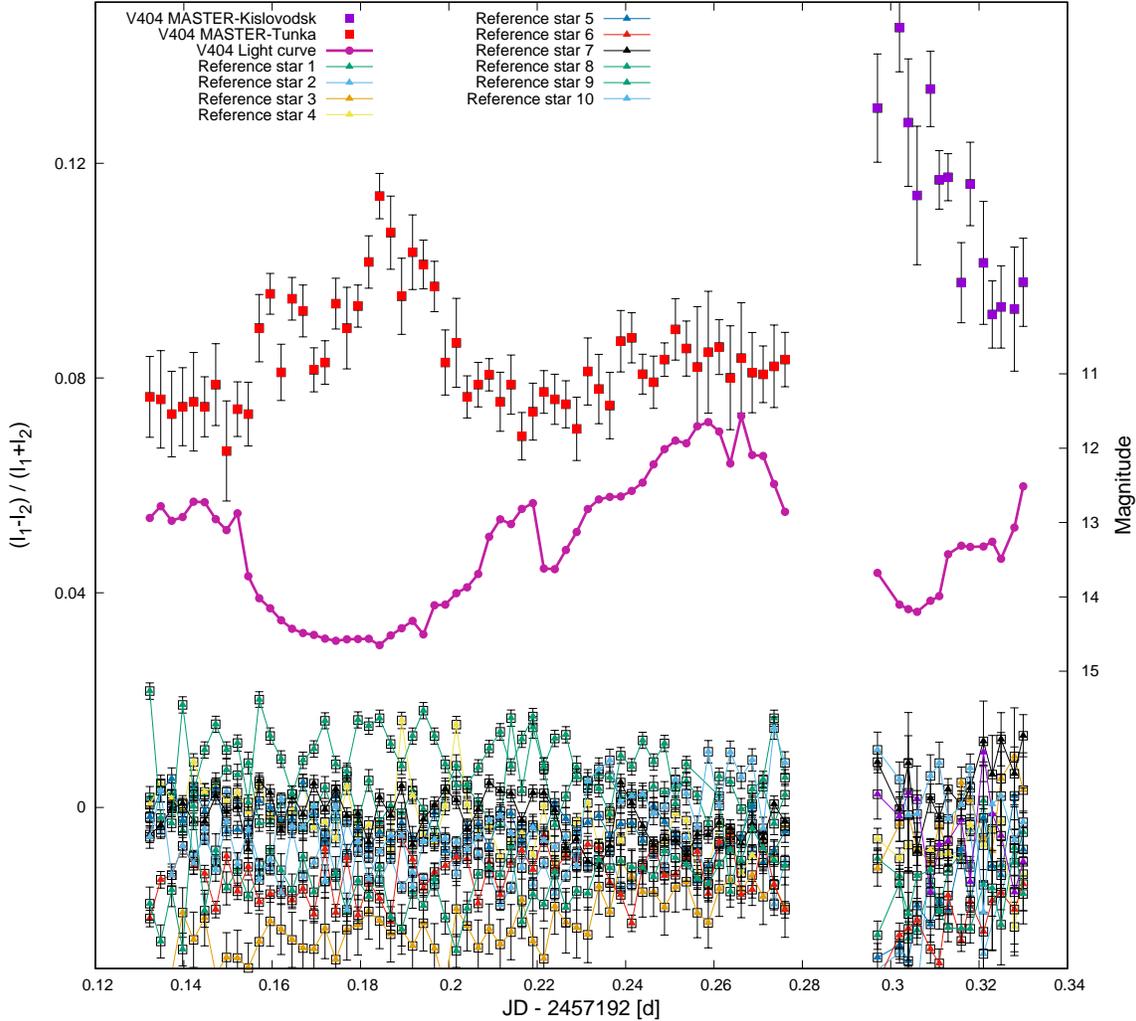}
\caption{
The variation of the  $Y=(I_{1}-I_{2})/(I_{1}+I_{2})$ value of  V404 Cyg microquasar based on MASTER-Tunka and MASTER-Kislovodsk observations.  The data  from MASTER-Tunka are up to 19:00 U.T,  and from MASTER-Kislovodsk, after 19:00 U.T. There polarization of many field stars are also included. The behaviour of V404 light curve at the time of polarization variability is demonstrated by the magenta curve.
}
\end{figure}

For the polarization measurements, we used the field stars with corresponding similar stellar magnitudes as the source, as follows. The error of the measurement is the standard deviation for the several measurements of the background stars. We use MASTER's wide fields (4 square degrees) to have measurements for thousands of field stars, many with magnitudes similar to V404 Cyg.



 We observed the Y value  
 to increase substantially,  up to 12\%, during V404 Cyg's fading.
The error was estimated from reference stars where we found the chi-square statistical significance of this event to be
99.99$\%$.

The second polarization variation event seen was recorded after receiving INTEGRAL alert N7035, on 2015 June. Twenty-seven seconds later, at 17:29:05UT, MASTER-Kislovodsk started photometric observations of V404 Cyg
in two mutually perpendicular polarizations parallel and perpendicular to the meridian (i.e.  $0^o$ and $90^o$.

After two hours MASTER recorded strong and fast flux increases, accompanied by a decrease of polarization from 14\% down to 8\%. This result
qualitatively corroborates the behavior found earlier in our MASTER-Tunka observations. The Tables of MASTER Y value  and photometry  data for V404 are available at http://master.sai.msu.ru/static/V404tables.pdf

\section{Interpretation and discussion} \label{sec:floats}

The main source of the optical emission in V404 Cyg  is the
$0.7^{+0.3}_{-0.2}M_\bigodot$ K3 III-type subgiant companion ~\citep{21,22,23}, including X-ray
radiation reprocessed to optical on the illuminated side, facing the black hole. The intrinsic  emission of the secondary (low-mass) star is
practically undetectable during the  time of the outburst, in which there is a factor of $\sim 10^8$, increase of the microquasar
luminosity. Thus the rise of the optical flux during this outburst is primarily due
to the heating of the low-mass star and, possibly, of the outer layers of the accretion disk. The accretion disk
forms as a result of the mass loss from the giant star, through the L1 point of its Roche lobe, into the gravitational potential well
 of the black
hole. While approaching the black hole the matter captured by it heats up to a temperature of several million degrees
and emits the observed X-ray radiation.

 However, there is yet another possible source of radiation in addition to the two mentioned above, namely a
 relativistic jet generated by the black hole and only observed until now in non-thermal radio and  hard X-ray emission. We
 believe that this  jet (or, more precisely, two symmetric bipolar jets) are responsible for the polarized optical emission.

 Here is our proposed scenario for the observed 2015 June 18 polarization variation in the microquasar:

\textbf{As the Y value is measured in two mutually perpendicular polarizers, its variability can be connected with variability of size of linear polarization or with turn of a vector of the  polarization plane.
 The first explanation seems more preferably as it is difficult to imagine an essential variability of jet orientation  on the times about tens of minutes.
The variability of flux in different polarizers can be most possibly explained by the variability of the contribution of strongly polarized light in the general flux of an optical emission. Not incidentally, the polarization increases when the general optical flux of the system falls and the contribution polarization component grows. Anti-correlation between the value of polarization and brightness of an optical flux  is apparently connected with the increase of jet contribution  during  the decreasing of the effect of warming up of an optical companion by x-ray emission  arising near the black hole.}

 At 15:36:00 (UT) the accreting
 mass inflow began to decrease rapidly near the black hole, the region of maximum energy release. The X-ray flux
 also decreased and, as a result, so did the intensity of the heating of the secondary (giant) star. At the same time the active
 emission of the jet continued, because it is not directly associated with the instantaneous accretion rate but
 rather with the acceleration of relativistic particles near the black hole. In other words, the jet became visible
 because its flux contribution increased with respect to the decrease of the luminosity of the irradiated secondary star. The jet emission is due to
 non-thermal synchrotron radiation and is therefore polarized, and hence its increased contribution resulted in the
 increase of the polarization of the total emission.

  Strong linear polarization (up to $\sim 30\%$) is observed in the jets of  BL Lac type objects, or blazars,  namely  quasars with
  relativistic jets pointing in the direction of the Earth. This is due to the synchrotron mechanism of their
  emission. Thus the discovery of non-thermal polarized optical component provides the direct evidence  that the
  nature of one of the most secure black hole candidates, namely V404 Cyg, is similar to that of the millions of times more massive
  quasars (blazars). If we assume that the total jet polarization is equal to $\simeq$30 $\%$, as observed in blazar type objects, then
  the jets in V404 Cyg contribute about K = $L_{j} /L_{o}$ $  \sim 15\%$ to the total luminosity, implying a jet
  optical flux of about $2\cdot10^{-11}$ ${\rm erg} {\rm s}^{-1} {\rm cm}^{-2}$. For the adopted distance of 2.39 $\pm$ 0.14 kpc ~\citep{24} we
  obtain a total jet luminosity of $\sim$ 4 $\cdot 10^{34}$ erg/s. The detection of unprecedentedly powerful
  non-thermal radio emission amounting to 6 Jy during outbursts ~\citep{14} provides independent evidence confirming
  the likely existence of a non-thermal jet during this last outburst. This fact agrees quite well with jet flux estimates inferred in
  terms of the synchrotron emission mechanism.

During this paper's preparation, information about independent polarimetry observations was published \citep{20}, where the authors wrote that they didn't see polarization variability. We stress that we have a more extensive data set and also don't see polarization variability, except unequivocally at two diferent times, during rapid and deep optical flux decreases, which is the subject of the paper. The optical flux decrease was over a short timescale (compared to the scale of the several dozens days in total of the MASTER monitoring observations) and has the character of a rapid decrease of the optical flux of the system. The relative rarity of such events explains why were no other polarization variations have been seen in V404 Cyg (e.g. by \cite{20}).

\section*{Acknowledgments}
MASTER Global Robotic Net is supported in part by the  Development Programm of Lomonosov Moscow State University.
This work was also supported in part by RFBR 15-02-07875 grant, Russian Science Foundation 16-12-00085 and National Research Foundation of South Africa.
We are grateful to Professor Tanaka for the information in his astro-ph paper, prior to publication. We are gratefull to referee for the number of remarks and suggestions that have improved the paper.





\pagestyle{empty}
\begin{table}[h]
\caption {Photometry of V404 Cyg by MASTER-Tunka Telescope in JD2457192.13-JD2457192.28 (EAST and WEST MASTER tubes)}
\begin{turn}{90}
\tiny
\begin{tabular}{|c|c|c|c|c|c|c|c|c|c|c|c|c|c|c|c|c|c|c|c|c|c|c|c|c|c|}
\hline
   JD    & Filter & Tube  & \#1      & \#2      & \#3      & \#4      & \#5      & \#6      & \#7      & \#8      & \#9      & \#10     & \#11     & \#12     & \#13     & \#14     & \#15     & \#16     & \#17     & \#18    &\#19 \\
\hline
    2457192.1323 & P     & EAST  & 12.855 & 12.623 & 12.267 & 12.479 & 12.252 & 12.858 & 12.820 & 12.779 & 12.305 & 12.713 & 13.737 & 13.124 & 14.408 & 13.074 & 13.344 & 14.804 & 14.335 & 13.368 & 14.888 \\
    2457192.1348 & P     & EAST  & 12.698 & 12.679 & 12.260 & 12.476 & 12.245 & 12.848 & 12.819 & 12.760 & 12.307 & 12.742 & 13.798 & 13.147 & 14.334 & 13.073 & 13.326 & 14.842 & 14.327 & 13.380 & 14.834 \\
    2457192.1372 & P     & EAST  & 12.897 & 12.681 & 12.259 & 12.472 & 12.257 & 12.849 & 12.823 & 12.760 & 12.304 & 12.729 & 13.794 & 13.130 & 14.398 & 13.085 & 13.325 & 14.809 & 14.361 & 13.384 & 14.828 \\
    2457192.1397 & P     & EAST  & 12.845 & 12.687 & 12.263 & 12.473 & 12.266 & 12.856 & 12.831 & 12.752 & 12.303 & 12.716 & 13.757 & 13.158 & 14.401 & 13.076 & 13.315 & 14.810 & 14.367 & 13.373 & 14.775 \\
    2457192.1422 & P     & EAST  & 12.638 & 12.694 & 12.263 & 12.494 & 12.250 & 12.858 & 12.813 & 12.760 & 12.318 & 12.720 & 13.778 & 13.110 & 14.388 & 13.078 & 13.323 & 14.806 & 14.357 & 13.366 & 14.843 \\
    2457192.1447 & P     & EAST  & 12.646 & 12.689 & 12.257 & 12.488 & 12.253 & 12.860 & 12.822 & 12.759 & 12.313 & 12.720 & 13.764 & 13.132 & 14.398 & 13.071 & 13.310 & 14.812 & 14.346 & 13.385 & 14.859 \\
    2457192.1472 & P     & EAST  & 12.870 & 12.683 & 12.265 & 12.492 & 12.253 & 12.851 & 12.813 & 12.771 & 12.307 & 12.713 & 13.766 & 13.126 & 14.374 & 13.079 & 13.330 & 14.766 & 14.348 & 13.367 & 14.796 \\
\end{tabular}
\end{turn}
\end{table}
\newpage
\pagestyle{empty}
\begin{table}[h]
\begin{turn}{90}
\tiny
 \begin{tabular}{|c|c|c|c|c|c|c|c|c|c|c|c|c|c|c|c|c|c|c|c|c|c|c|c|c|c|}
   JD    & Filter & Tube  & \#1      & \#2      & \#3      & \#4      & \#5      & \#6      & \#7      & \#8      & \#9      & \#10     & \#11     & \#12     & \#13     & \#14     & \#15     & \#16     & \#17     & \#18    &\#19 \\
\hline
    2457192.2414 & P     & EAST  & 12.690 & 12.733 & 12.266 & 12.475 & 12.269 & 12.885 & 12.817 & 12.740 & 12.341 & 12.727 & 13.790 & 13.138 & 14.414 & 13.073 & 13.293 & 14.847 & 14.334 & 13.354 & 14.809 \\
    2457192.2438 & P     & EAST  & 12.451 & 12.722 & 12.265 & 12.471 & 12.273 & 12.890 & 12.820 & 12.737 & 12.324 & 12.728 & 13.799 & 13.137 & 14.407 & 13.072 & 13.298 & 14.844 & 14.355 & 13.358 & 14.790 \\
    2457192.2463 & P     & EAST  & 12.244 & 12.730 & 12.265 & 12.482 & 12.260 & 12.871 & 12.817 & 12.747 & 12.341 & 12.728 & 13.792 & 13.133 & 14.416 & 13.075 & 13.288 & 14.842 & 14.351 & 13.359 & 14.831 \\
    2457192.2488 & P     & EAST  & 12.127 & 12.723 & 12.261 & 12.475 & 12.257 & 12.887 & 12.820 & 12.748 & 12.335 & 12.734 & 13.799 & 13.135 & 14.424 & 13.072 & 13.294 & 14.871 & 14.339 & 13.355 & 14.781 \\
    2457192.2513 & P     & EAST  & 12.166 & 12.733 & 12.265 & 12.469 & 12.265 & 12.882 & 12.828 & 12.738 & 12.329 & 12.731 & 13.789 & 13.150 & 14.415 & 13.075 & 13.288 & 14.858 & 14.351 & 13.358 & 14.825 \\
    2457192.2538 & P     & EAST  & 11.936 & 12.732 & 12.264 & 12.468 & 12.275 & 12.879 & 12.828 & 12.734 & 12.328 & 12.727 & 13.637 & 13.138 & 14.419 & 13.076 & 13.298 & 14.837 & 14.346 & 13.364 & 14.818 \\
    2457192.2562 & P     & EAST  & 11.875 & 12.720 & 12.273 & 12.479 & 12.275 & 12.870 & 12.816 & 12.741 & 12.341 & 12.722 & 13.669 & 13.134 & 14.433 & 13.070 & 13.296 & 14.851 & 14.352 & 13.355 & 14.826 \\
\end{tabular}
\end{turn}
\end{table}
\newpage
\pagestyle{empty}
\begin{table}[h]
\begin{turn}{90}
\tiny
 \begin{tabular}{|c|c|c|c|c|c|c|c|c|c|c|c|c|c|c|c|c|c|c|c|c|c|c|c|c|c|}
   JD    & Filter & Tube  & \#1      & \#2      & \#3      & \#4      & \#5      & \#6      & \#7      & \#8      & \#9      & \#10     & \#11     & \#12     & \#13     & \#14     & \#15     & \#16     & \#17     & \#18    &\#19 \\
\hline
    2457192.1323 & P     & WEST  & 13.021 & 12.752 & 12.238 & 12.442 & 12.220 & 12.871 & 12.783 & 12.701 & 12.280 & 12.669 & 13.751 & 13.052 & 14.544 & 13.029 & 13.225 & 14.955 & 14.404 & 13.319 & 14.771 \\
    2457192.1348 & P     & WEST  & 12.863 & 12.750 & 12.226 & 12.425 & 12.214 & 12.866 & 12.785 & 12.690 & 12.273 & 12.694 & 13.753 & 13.052 & 14.582 & 13.039 & 13.200 & 15.012 & 14.388 & 13.284 & 14.778 \\
    2457192.1372 & P     & WEST  & 13.056 & 12.748 & 12.224 & 12.432 & 12.220 & 12.863 & 12.794 & 12.692 & 12.282 & 12.688 & 13.747 & 13.056 & 14.547 & 13.017 & 13.217 & 15.066 & 14.407 & 13.281 & 14.765 \\
    2457192.1397 & P     & WEST  & 13.007 & 12.740 & 12.219 & 12.440 & 12.231 & 12.854 & 12.791 & 12.702 & 12.279 & 12.683 & 13.764 & 13.066 & 14.571 & 13.026 & 13.238 & 14.953 & 14.402 & 13.300 & 14.738 \\
    2457192.1422 & P     & WEST  & 12.802 & 12.759 & 12.228 & 12.435 & 12.229 & 12.854 & 12.782 & 12.697 & 12.290 & 12.675 & 13.745 & 13.071 & 14.565 & 13.016 & 13.230 & 15.000 & 14.369 & 13.291 & 14.774 \\
    2457192.1447 & P     & WEST  & 12.808 & 12.744 & 12.226 & 12.439 & 12.223 & 12.865 & 12.784 & 12.696 & 12.287 & 12.678 & 13.751 & 13.054 & 14.552 & 13.026 & 13.229 & 15.000 & 14.376 & 13.305 & 14.745 \\
    2457192.1472 & P     & WEST  & 13.041 & 12.745 & 12.239 & 12.435 & 12.226 & 12.859 & 12.785 & 12.694 & 12.289 & 12.682 & 13.764 & 13.073 & 14.561 & 13.018 & 13.227 & 14.999 & 14.377 & 13.299 & 14.730 \\
\end{tabular}
\end{turn}
\end{table}
\newpage
\pagestyle{empty}
\begin{table}[h]
\begin{turn}{90}
\tiny
 \begin{tabular}{|c|c|c|c|c|c|c|c|c|c|c|c|c|c|c|c|c|c|c|c|c|c|c|c|c|c|}
   JD    & Filter & Tube  & \#1      & \#2      & \#3      & \#4      & \#5      & \#6      & \#7      & \#8      & \#9      & \#10     & \#11     & \#12     & \#13     & \#14     & \#15     & \#16     & \#17     & \#18    &\#19 \\
\hline
 2457192.2438 & P     & WEST  & 12.600 & 12.745 & 12.235 & 12.425 & 12.213 & 12.870 & 12.777 & 12.680 & 12.303 & 12.687 & 13.781 & 13.073 & 14.597 & 13.024 & 13.216 & 15.045 & 14.438 & 13.289 & 14.751 \\
    2457192.2463 & P     & WEST  & 12.409 & 12.742 & 12.243 & 12.422 & 12.216 & 12.858 & 12.787 & 12.681 & 12.308 & 12.693 & 13.782 & 13.074 & 14.581 & 13.025 & 13.223 & 15.038 & 14.420 & 13.290 & 14.755 \\
    2457192.2488 & P     & WEST  & 12.301 & 12.745 & 12.235 & 12.427 & 12.221 & 12.863 & 12.774 & 12.677 & 12.303 & 12.690 & 13.774 & 13.075 & 14.611 & 13.023 & 13.219 & 15.045 & 14.414 & 13.289 & 14.752 \\
    2457192.2513 & P     & WEST  & 12.334 & 12.754 & 12.243 & 12.423 & 12.222 & 12.864 & 12.778 & 12.674 & 12.292 & 12.688 & 13.783 & 13.070 & 14.613 & 13.030 & 13.214 & 15.030 & 14.424 & 13.294 & 14.762 \\
    2457192.2538 & P     & WEST  & 12.102 & 12.747 & 12.235 & 12.423 & 12.217 & 12.857 & 12.770 & 12.680 & 12.297 & 12.694 & 13.783 & 13.071 & 14.599 & 13.032 & 13.212 & 15.037 & 14.423 & 13.289 & 14.772 \\
    2457192.2562 & P     & WEST  & indef & 12.741 & 12.243 & 12.429 & 12.210 & 12.857 & 12.772 & 12.684 & 12.303 & 12.699 & 13.776 & 13.066 & 14.586 & 13.025 & 13.217 & 15.036 & 14.426 & 13.284 & 14.750 \\
    2457192.2587 & P     & WEST  & 12.172 & 12.747 & 12.242 & 12.423 & 12.217 & 12.849 & 12.783 & 12.675 & 12.311 & 12.694 & 13.776 & 13.082 & 14.588 & 13.014 & 13.215 & 15.061 & 14.417 & 13.286 & 14.773 \\
\hline
    \end{tabular}
\end{turn}
\end{table}

\newpage
\pagestyle{empty}
\begin{table}[h]
\caption{$(I_{West}-I_{East})/(I_{West}+{I_{East}})$ of V404 Cyg by MASTER-Tunka Telescope in JD2457192.13-JD2457192.28}
\begin{turn}{90}
\tiny
 \begin{tabular}{|c|c|c|c|c|c|c|c|c|c|c|c|c|c|c|c|c|c|c|c|c|c|c|c|c|c|}
\hline
   JD    & \#1   & \#2   & \#3   & \#4   & \#5   & \#6   & \#7   & \#8   & \#9   & \#10  & \#11  & \#12  & \#13  & \#14  & \#15  & \#16  & \#17  & \#18 &\#19\\
    \hline
    2457192.1323 & 0.07648 & 0.05952 & -0.01317 & -0.01685 & -0.01455 & 0.00617 & -0.01685 & -0.03572 & -0.01133 & -0.02008 & 0.00663 & -0.03296 & 0.06273 & -0.02054 & -0.05456 & 0.06961 & 0.03195 & -0.02238 & -0.05364 \\
    2457192.1348 & 0.07602 & 0.03287 & -0.01547 & -0.02330 & -0.01409 & 0.00847 & -0.01547 & -0.03204 & -0.01547 & -0.02192 & -0.02054 & -0.04354 & 0.11390 & -0.01547 & -0.05778 & 0.07831 & 0.02827 & -0.04400 & -0.02560 \\
    2457192.1372 & 0.07327 & 0.03103 & -0.01593 & -0.01823 & -0.01685 & 0.00663 & -0.01317 & -0.03112 & -0.00995 & -0.01869 & -0.02146 & -0.03388 & 0.06869 & -0.03112 & -0.04951 & 0.11798 & 0.02136 & -0.04721 & -0.02882 \\
    2457192.1397 & 0.07465 & 0.02459 & -0.02008 & -0.01501 & -0.01593 & -0.00074 & -0.01823 & -0.02284 & -0.01087 & -0.01501 & 0.00341 & -0.04216 & 0.07831 & -0.02284 & -0.03526 & 0.06594 & 0.01630 & -0.03342 & -0.01685 \\
    2457192.1422 & 0.07556 & 0.03011 & -0.01593 & -0.02698 & -0.00949 & -0.00166 & -0.01409 & -0.02882 & -0.01271 & -0.02054 & -0.01501 & -0.01777 & 0.08151 & -0.02836 & -0.04262 & 0.08929 & 0.00571 & -0.03434 & -0.03158 \\
    2457192.1447 & 0.07465 & 0.02551 & -0.01409 & -0.02238 & -0.01363 & 0.00249 & -0.01731 & -0.02882 & -0.01179 & -0.01916 & -0.00580 & -0.03572 & 0.07098 & -0.02054 & -0.03710 & 0.08654 & 0.01400 & -0.03664 & -0.05227 \\
    2457192.1472 & 0.07877 & 0.02873 & -0.01179 & -0.02606 & -0.01225 & 0.00387 & -0.01271 & -0.03526 & -0.00810 & -0.01409 & -0.00074 & -0.02422 & 0.08609 & -0.02790 & -0.04721 & 0.10707 & 0.01354 & -0.03112 & -0.03020 \\
\hline
    \end{tabular}
\end{turn}
\end{table}
\tiny

    \newpage
\pagestyle{empty}
\begin{table}[h]
\tiny
\begin{turn}{90}
 \begin{tabular}{|c|c|c|c|c|c|c|c|c|c|c|c|c|c|c|c|c|c|c|c|c|c|c|c|c|c|}
\hline
   JD    & \#1   & \#2   & \#3   & \#4   & \#5   & \#6   & \#7   & \#8   & \#9   & \#10  & \#11  & \#12  & \#13  & \#14  & \#15  & \#16  & \#17  & \#18 &\#19\\
    \hline
    2457192.2687 & 0.07556 & 0.00709 & -0.01133 & -0.02514 & -0.02054 & -0.01087 & -0.01777 & -0.02882 & -0.00995 & -0.01501 & -0.00903 & -0.02836 & 0.07694 & -0.02238 & -0.03250 & 0.08883 & 0.03839 & -0.03296 & -0.03020 \\
    2457192.2711 & 0.07419 & 0.00249 & -0.01501 & -0.02882 & -0.02606 & 0.00018 & -0.01593 & -0.02698 & -0.02008 & -0.00995 & 0.00295 & -0.02882 & 0.08151 & -0.02146 & -0.03434 & 0.08563 & 0.03379 & -0.02974 & -0.03020 \\
    2457192.2736 & 0.08014 & 0.00847 & -0.01455 & -0.02422 & -0.01823 & -0.00903 & -0.01962 & -0.03158 & -0.01179 & -0.01547 & -0.00672 & -0.02422 & 0.09020 & -0.02468 & -0.03572 & 0.07282 & 0.03793 & -0.02836 & -0.01823 \\
    2457192.2761 & 0.07282 & 0.00985 & -0.01593 & -0.02376 & -0.02606 & -0.00212 & -0.01593 & -0.02606 & -0.00949 & -0.01501 & -0.00764 & -0.02836 & 0.07831 & -0.01823 & -0.03710 & 0.06778 & 0.03057 & -0.03158 & -0.01639 \\
    2457192.2786 & 0.07053 & 0.00018 & -0.01409 & -0.03112 & -0.02514 & -0.00442 & -0.01685 & -0.02468 & -0.01823 & -0.01133 & -0.00120 & -0.02698 & 0.07556 & -0.01916 & -0.04170 & 0.09294 & 0.04666 & -0.02146 & -0.01823 \\
\hline
    \end{tabular}
\end{turn}
\end{table}
\tiny
\newpage

\pagestyle{empty}
\begin{table}[h]
\tiny
\caption {Photometry of V404 Cyg by MASTER-Kislovodsk Telescope in JD2457192.13-JD2457192.28}
\begin{turn}{90}
\tiny
 \begin{tabular}{|c|c|c|c|c|c|c|c|c|c|c|c|c|c|c|c|c|c|c|c|c|c|c|c|c|c|}
\hline
   JD    & Filter & Tube  & \#1      & \#2      & \#3      & \#4      & \#5      & \#6      & \#7      & \#8      & \#9      & \#10     & \#11     & \#12     & \#13     & \#14     & \#15     & \#16     & \#17     & \#18    &\#19 \\
\hline
    2457192.3020 & P     & EAST  & 13.940 & 12.662 & 12.260 & 12.463 & 12.235 & 12.866 & 12.803 & 12.724 & 12.306 & 12.696 & 13.788 & 13.110 & 14.422 & 13.038 & 13.260 & 14.831 & 14.316 & 13.326 & 14.793 \\
    2457192.3040 & P     & EAST  & 14.014 & 12.638 & 12.262 & 12.462 & 12.239 & 12.863 & 12.810 & 12.718 & 12.308 & 12.698 & 13.784 & 13.108 & 14.440 & 13.036 & 13.248 & 14.860 & 14.316 & 13.328 & 14.787 \\
    2457192.3060 & P     & EAST  & 14.064 & 12.626 & 12.248 & 12.460 & 12.238 & 12.854 & 12.800 & 12.731 & 12.297 & 12.701 & 13.787 & 13.114 & 14.459 & 13.041 & 13.254 & 14.850 & 14.313 & 13.335 & 14.810 \\
    2457192.3090 & P     & EAST  & 13.918 & 12.613 & 12.254 & 12.466 & 12.241 & 12.868 & 12.795 & 12.724 & 12.303 & 12.694 & 13.793 & 13.108 & 14.435 & 13.032 & 13.250 & 14.847 & 14.329 & 13.329 & 14.829 \\
    2457192.3110 & P     & EAST  & 13.863 & 12.477 & 12.258 & 12.455 & 12.239 & 12.859 & 12.800 & 12.730 & 12.300 & 12.703 & 13.803 & 13.109 & 14.447 & 13.041 & 13.254 & 14.862 & 14.329 & 13.331 & 14.806 \\
    2457192.3130 & P     & EAST  & 13.302 & 12.538 & 12.260 & 12.466 & 12.244 & 12.853 & 12.789 & 12.730 & 12.292 & 12.701 & 13.794 & 13.118 & 14.452 & 13.046 & 13.247 & 14.814 & 14.328 & 13.326 & 14.834 \\
    2457192.3160 & P     & EAST  & 13.203 & 12.466 & 12.252 & 12.462 & 12.242 & 12.845 & 12.796 & 12.745 & 12.285 & 12.693 & 13.811 & 13.125 & 14.411 & 13.048 & 13.248 & 14.864 & 14.353 & 13.325 & 14.808 \\
\hline
    \end{tabular}
    \end{turn}
\end{table}

\newpage
\pagestyle{empty}
\tiny
\begin{table}[h]
\caption {$(I_{West}-I_{East})/(I_{West}+{I_{East}})$ of V404 Cyg by MASTER-Kislovodsk Telescope in JD2457192.13-JD2457192.28}
\tiny
\begin{turn}{90}
 \begin{tabular}{|c|c|c|c|c|c|c|c|c|c|c|c|c|c|c|c|c|c|c|c|c|c|c|c|c|c|}
\hline
   JD    & \#1   & \#2   & \#3   & \#4   & \#5   & \#6   & \#7   & \#8   & \#9   & \#10  & \#11  & \#12  & \#13  & \#14  & \#15  & \#16  & \#17  & \#18 &\#19\\
    \hline
    2457192.2970 & 0.14225 & 0.03682 & 0.00000 & -0.02072 & 0.00368 & 0.02992 & -0.01842 & -0.02302 & 0.00092 & 0.01796 & 0.01197 & -0.01428 & 0.13638 & 0.02026 & -0.00184 & 0.13819 & 0.10871 & -0.00322 & -0.00184 \\
    2457192.3020 & 0.15081 & 0.04280 & -0.00553 & -0.01980 & -0.00230 & 0.01934 & -0.01842 & -0.01658 & 0.01335 & 0.00691 & 0.00553 & -0.00691 & 0.13095 & 0.01243 & 0.00414 & 0.16204 & 0.10780 & 0.00461 & 0.00368 \\
    2457192.3040 & 0.13909 & 0.04924 & -0.01566 & -0.02210 & 0.00184 & 0.02302 & -0.01980 & -0.01381 & 0.00737 & 0.01704 & 0.01151 & -0.01105 & 0.12280 & 0.00737 & 0.00737 & 0.15261 & 0.11962 & -0.00461 & 0.01105 \\
    2457192.3060 & 0.12552 & 0.05199 & -0.01289 & -0.02486 & 0.00184 & 0.02946 & -0.00829 & -0.01105 & 0.01013 & 0.00184 & 0.01151 & -0.00276 & 0.10780 & 0.00875 & 0.00230 & 0.14225 & 0.12370 & -0.01612 & -0.00921 \\
    2457192.3090 & 0.12370 & 0.07080 & -0.00829 & -0.00276 & -0.00230 & 0.02256 & -0.01612 & -0.02072 & 0.00230 & 0.00737 & -0.01013 & -0.00599 & 0.11917 & 0.01151 & -0.00461 & 0.12869 & 0.09732 & 0.00000 & -0.01842 \\
    2457192.3110 & 0.11599 & 0.13231 & -0.01289 & -0.00276 & -0.00276 & 0.02946 & -0.01151 & -0.02302 & 0.00322 & 0.00414 & -0.00092 & -0.01335 & 0.10598 & 0.01428 & 0.00276 & 0.11826 & 0.08865 & -0.00368 & -0.02072 \\
    2457192.3130 & 0.11371 & 0.10689 & -0.00921 & -0.00599 & -0.00599 & 0.02578 & -0.01658 & -0.01381 & 0.01612 & 0.00599 & -0.00368 & -0.01980 & 0.11462 & 0.00046 & -0.00599 & 0.14766 & 0.07996 & 0.00184 & -0.01013 \\
\hline
    \end{tabular}
\end{turn}
\end{table}
\newpage



\begin{thebibliography}{}

 \bibitem[\protect\citeauthoryear{Antokhina et al.}{1993}]{BestBH} Antokhina et al., 1993, AZh, 70, 804A
\bibitem[\protect\citeauthoryear{Barthelmy et al.}{2015}]{7} Barthelmy S. et al. 2015,  GCN,  17929, 1
\bibitem[\protect\citeauthoryear{Barthelmy et al.}{1994}]{Barthelmy94} Barthelmy S. et al. 1994, AIP Conf. Proc., 307, 643
\bibitem[\protect\citeauthoryear{Barthelmy et al.}{1995}]{Barthelmy95} Barthelmy S. et al. 1995, Ap\&SS, 231, 235
\bibitem[\protect\citeauthoryear{Blay et al.}{2015}]{18} Blay et al. 2015, ATel, 7678, 1
\bibitem[\protect\citeauthoryear{Casares et al.}{1994}]{21} Casares J., Charles P. 1994, Nature, 355, 614
\bibitem[\protect\citeauthoryear{Casares et al.}{1991}]{casares} Casares J., Charles P. A., Jones, D. H. P., et al. 1991, MNRAS, 250, 712
\bibitem[\protect\citeauthoryear{Ferrigno et al.}{2015}]{12} Ferrigno et al. 2015, ATel 7662, 1
\bibitem[\protect\citeauthoryear{Gorbovskoy et al.}{2012}]{gor2012} Gorbovskoy E.  et al., 2012, MNRAS, 421, 1874
\bibitem[\protect\citeauthoryear{Gorbovskoy et al.}{2016}]{gor2016} Gorbovskoy E.  et al., 2016, MNRAS, 455, 3312
\bibitem[\protect\citeauthoryear{Itoh et al.}{2015}]{19}Itoh et al. 2015, ATel 7709, 1
\bibitem[\protect\citeauthoryear{Khargharia et al.}{2010}]{23} Khargharia et al. 2010, Astrophys.J., 716, 1105
\bibitem[\protect\citeauthoryear{Kimura et al.}{2016}]{999} Kimura M. et al. 2016, Nature, 529, 7584, pp. 54-58
\bibitem[\protect\citeauthoryear{Kornilov et al.}{2012}]{25} Kornilov et al., 2012,  2012, Experimental Astronomy, Vol. 33, Issue 1, pp.173-196, 2012
    \bibitem[\protect\citeauthoryear{Lipunov et al.}{2007}]{lipunov2007}Lipunov V.~M. et~al., 2007, Astronomy Reports, 51, 1004
\bibitem[\protect\citeauthoryear{Lipunov et al.}{2010}]{8} Lipunov V.M. et al. 2010, Advances in Astronomy, 349171
\bibitem[\protect\citeauthoryear{Lipunov et al.}{2015}]{9} Lipunov V.M.et al. 2015, ATel, 7696, 1
\bibitem[\protect\citeauthoryear{Lipunov et al.}{2016}]{Lipunov2016a} Lipunov et al., 2015, MNRAS, 455, 712
\bibitem[\protect\citeauthoryear{Makino et al.}{1989}]{1} Makino F. et al. 1989, IAUC  4782, 1
\bibitem[\protect\citeauthoryear{Miller et al.}{2009}]{24} Miller-Jones et al., 2009, Mon. Not. R. Astron. Soc., 394, 1440
\bibitem[\protect\citeauthoryear{Monet et al.}{2003}]{monet} Monet D. et al.,  2003 Astron. J., Vol. 125, p. 984
\bibitem[\protect\citeauthoryear{Mooley et al.}{2015}]{10} Mooley K. et al., 2015, ATel 7658, 1
\bibitem[\protect\citeauthoryear{Motta et al.}{2015}]{13} Motta et al. 2015, ATel 7666, 1
\bibitem[\protect\citeauthoryear{Munoz-Darias et al.}{2015}]{11} Munoz-Darias et al. 2015, ATel 7659, 1
\bibitem[\protect\citeauthoryear{Pruzhinskaya et al.}{2014}]{17}Pruzhinskaya et al., 2014, New Astronomy, 29, 65
\bibitem[\protect\citeauthoryear{Rana et al.}{2015}]{6} Rana et al. 2015, Astrophys.J. in press, http://arxiv.org/abs/1507.04049
\bibitem[\protect\citeauthoryear{Rodriguez et al.}{2015}]{4} Rodriguez et al. 2015, Astron. $\&$ Astrophys J.Let. in press, http://arxiv.org/abs/1507.06659
\bibitem[\protect\citeauthoryear{Shahbaz et al.}{1994}]{22} Shahbaz et al. 1994, Mon. Not. R. Astron. Soc.  271, L10
\bibitem[\protect\citeauthoryear{Tanaka et al.}{2016}]{20}Tanaka et al. 2016, ApJ, 823, 35
\bibitem[\protect\citeauthoryear{Tetarenko et al.}{2015a}]{14} Tetarenko et al. 2015, ATel 7661, 1
\bibitem[\protect\citeauthoryear{Tetarenko et al.}{2015b}]{15} Tetarenko et al. 2015, ATel 7708, 1
\bibitem[\protect\citeauthoryear{Tsubono et al.}{2015}]{16} Tsubono et al. 2015, ATel 7701, 1
\bibitem[\protect\citeauthoryear{Wagner et al.}{1989a}]{2} Wagner R.M. et al. 1989a, IAUC 4783, 1
\bibitem[\protect\citeauthoryear{Wagner et al.}{1989b}]{3} Wagner R.M. et al. 1989b, IAUC 4797, 1

\end{thebibliography}
\end{document}